\PassOptionsToPackage{table}{xcolor}
\documentclass[manuscript,screen]{acmart}
\AtBeginDocument{%
  }

\setcopyright{acmlicensed}
\copyrightyear{2018}
\acmYear{2025}
\acmDOI{XXXXXXX.XXXXXXX}

\acmISBN{978-1-4503-XXXX-X/18/06}

\usepackage{array}
\usepackage{multirow}

\usepackage{amsmath, amsthm, geometry, fancyhdr, hyperref}
\usepackage{graphicx}
\usepackage{textcomp}
\usepackage{geometry}
\usepackage{float}
\usepackage{algorithm}
\usepackage[noend]{algorithmic}
\usepackage{adjustbox}
\begin{document}

\title{Algorithms for estimating linear function in data mining}

\author{Thomas Hoang}
\email{hoang_t2@denison.edu}
\affiliation{%
  \institution{Denison University}
  \city{Granville}
  \state{Ohio}
  \country{USA}
}










\begin{abstract}
    The main goal of this topic is to showcase several studied algorithms for estimating the linear utility function to predict the users' preferences. For example, if a user comes to buy a car that has several attributes including speed, color, age, etc in a linear function, the algorithms that we present in this paper help with estimating this linear function to filter out a small subset that would be of best interest to the user among a million tuples in a very large database. In addition, the estimating linear function could also be applicable in getting to know what the data can do or predicting the future based on the data that is used in data science, which is demonstrated by the GNN, PLOD \cite{Hoang2024GNNGN} \cite{hoang2024plod} algorithms. In the ever-evolving field of data science, deriving valuable insights from large datasets is critical for informed decision-making, particularly in predictive applications. Data analysts often identify high-quality datasets without missing values, duplicates, or inconsistencies before merging diverse attributes for analysis. Taking housing price prediction as a case study, various attributes must be considered, including location factors (proximity to urban centers, crime rates), property features (size, style, modernity), and regional policies (tax implications). Experts in the field typically rank these attributes to establish a predictive utility function, which machine learning models use to forecast outcomes like housing prices. Several data discovery algorithms, including \cite{Hoang2024GNNGN}, \cite{hoang2024plod}, \cite{Hoang2024BODBO} address the challenges of predefined utility functions and human input for attribute ranking, which often result in a time-consuming iterative process, which the work of \cite{Galhotra2023MetamGD} cannot overcome. The notable enhancement uses a Graph Neural Network (GNN) algorithm that builds on previous approaches. The GNN algorithm leverages the power of graph neural networks and large language models to interpret text-based values that earlier models like PLOD could not handle, significantly improving the reliability of outcome predictions. GNN extends PLOD's capabilities by incorporating numerical and textual data, offering a comprehensive approach to understanding user preferences for data science and analytics applications.
\end{abstract}





\maketitle

\section{Introduction}

The method of indistinguishability query \cite{Lall2024TheIQ} for identifying tuples that are close to optimal from a user's perspective. This query retrieves all tuples that are only a small margin below the optimal value according to the user's unknown utility function. This approach is based on the insight that users often struggle to differentiate between very similar options, and even slightly suboptimal tuples may possess attributes that make them appealing. In addition, this framework asks the user to make a limited number of comparisons, helping to refine an understanding of their preferences. This is a big advancement compared to traditional that ask users for utility functions to proceed with filtering out user's preference of tuples, like Top-K algorithms \cite{QY12}, \cite{SI07},  \cite{Khosla2015TopkQP}, and \cite{Xiao2016ProbabilisticTR}, in which the Top-K algorithm requires the user's utility function, which can be difficult to obtain and is not always applicable where user preferences are not well-defined. 

In addition to estimating the user's preference for unknown utility functions. The estimating linear function could also be applicable in getting to know what the data can do or predicting the future based on the data that is used in data science. For instance, a scientist may prioritize specific attributes over others based on their domain expertise. For example, an apartment in a central area like New York City with low crime rates and modern amenities may be favored over a remote historic house. Even within similar urban environments, decision-making can be complex—such as weighing the pros and cons of a centrally located apartment against a slightly cheaper option in the suburbs. Other considerations, such as neighborhood friendliness or environmental tranquility, might influence rankings. A utility function captures these variations by assigning importance to each attribute in the prediction process.

Many machine learning algorithms and applied machine learning approaches \cite{Galhotra2023MetamGD} \cite{Fernandez2018AurumAD} \cite{Gong2021VerVD} often require predefined utility functions, meaning the user already understands the weights for each coefficient in linear function (supposedly ranging from 0 to 1), which can be impractical or difficult for many users to specify. However, these methods have limitations, as they may still produce data subsets that don't fully align with the user's actual utility function due to mismatches between the predicted and true utility functions.

Inspired by the approach of Indistinguishability Query, the approach, GNN (Graph Neural Networks and Large Language Models for Data Discovery), overcomes these challenges to predict the future. GNN leverages user-defined attribute rankings and incorporates advanced machine-learning techniques. By combining PLOD's numerical prediction capabilities with the power of Graph Neural Networks (GNNs) and Large Language Models (LLMs) for text-based data, GNN provides more accurate utility function estimations. This allows for selecting data subsets that more closely reflect user preferences and improves prediction outcomes. By integrating complex models capable of handling large, mixed-type datasets, GNN represents a significant advancement in multimodal predictive applications.

\section{Problem Definition}
\begin{table}[htbp]
  \caption{Notation and Meaning for GNN Algorithm}
  \label{tab:GNN_notation}
  \centering
  \begin{tabular}{|c|l|}
  \hline
  \textbf{Notation} & \textbf{Meaning} \\ 
  \hline
  $T$ & The set of all tuples \\ 
  $D$ & The dataset with both numerical and textual features \\ 
  $X$ & The set of all attributes (numerical and textual) in $D$ \\ 
  $X^{\text{num}}$ & The set of numerical attributes in $D$ \\ 
  $X^{\text{text}}$ & The set of textual attributes in $D$ \\ 
  $x_j^{\text{num}}$ & The $j$th numerical attribute in $X^{\text{num}}$ \\ 
  $x_k^{\text{text}}$ & The $k$th textual attribute in $X^{\text{text}}$ \\ 
  $\beta_{j,\text{num}}$ & Coefficient for the $j$th numerical attribute \\ 
  $\beta_{k,\text{text}}$ & Coefficient for the $k$th textual attribute \\ 
  $h_k^{\text{text}}$ & Feature embedding for the $k$th textual attribute generated by LLM \\ 
  $h_k^{\text{GNN}}$ & Combined feature embeddings processed by GNN \\ 
  $u(x)$ & Utility function estimating the score for a tuple \\ 
  $u_{\text{syn}}(x)$ & Synthetic utility function based on initial coefficient estimates \\ 
  $u_{\text{real}}(x)$ & Final utility function used for optimal tuple selection \\ 
  $t$ & A subset of $T$ containing optimal tuples \\ 
  \hline
  \end{tabular}
\end{table}

\textbf{Linear function}: 
\begin{itemize}
    \item $LINEAR = \{f|f(x) =  \sum\limits_{i=1}^{d}f(x_i) \\ \text{\{where each} \ f(x_i)  \text{ is a linear function.}\}$
\end{itemize}

\[
u(x) = a \cdot x_1 + b \cdot x_2 + c \cdot x_3 + \cdots + z \cdot x_d,
\]
\[
\text{where } u(x) \text{ is the utility function, and each term } f(x_i) \text{ is linear.}
\]

\subsection{Model Representation For GNN algorithm, adapted from GNN \cite{Hoang2024GNNGN}}

Below is the utility function:
\[
u(\mathbf{x}) = \sum_{j=1}^{m} \beta_{j,\text{num}} \cdot x_j^{\text{num}} + \sum_{k=1}^{n} \beta_{k,\text{text}} \cdot x_k^{\text{text}} + \epsilon
\]
where:
\begin{itemize}
    \item \( \mathbf{x} = (x_1^{\text{num}}, \ldots, x_m^{\text{num}}, x_1^{\text{text}}, \ldots, x_n^{\text{text}}) \) is the vector of attributes, with \( x_j^{\text{num}} \) representing numerical attributes and \( x_k^{\text{text}} \) representing textual attributes.
    \item \( \beta_{j,\text{num}} \) and \( \beta_{k,\text{text}} \) are the coefficients for numerical and textual attributes, respectively.
    \item \( \epsilon \) is the error term.
\end{itemize}

\subsection{Goal, adapted from GNN \cite{Hoang2024GNNGN}}

We want to estimate the coefficients \( \beta_{j,\text{num}} \) and \( \beta_{k,\text{text}} \) for the utility function \( u(\mathbf{x}) \), combining both numerical and textual data, to predict the utility score, \cite{Hoang2024GNNGN}.

\subsection{Numerical Data Processing, adapted from GNN \cite{Hoang2024GNNGN}}

For the numerical attributes, we want to apply linear regression techniques to estimate the coefficients \( \beta_{j,\text{num}} \). to minimize the sum of squared errors (SSE) between the observed utility scores and those predicted by the model, the model is trained as below:
\[
SSE_{\text{num}} = \sum_{i=1}^{N} (u_i - \hat{u}_i^{\text{num}})^2 = \sum_{i=1}^{N} \left( u_i - \sum_{j=1}^{m} \beta_{j,\text{num}} \cdot x_{i,j}^{\text{num}} \right)^2
\]
Where \( N \) is the number of data points. Adapted from \cite{hoang2024plod} as in estimating numerical data for the utility function

\subsection{Textual Data Processing, adapted from GNN \cite{Hoang2024GNNGN}}

For the textual attributes, we try to estimate the coefficients \( \beta_{k,\text{text}} \). Meanwhile, we let the GNN process the graph structure of the dataset, where each node represents an attribute, and each edge represents a relationship between attributes. LLM processes textual data and outputs feature embeddings that are fed into the GNN:
\[
\mathbf{h}_k^{\text{text}} = \text{GNN}(\text{LLM}(x_k^{\text{text}}))
\]
Our goal is to minimize the SSE for textual attributes:
\[
SSE_{\text{text}} = \sum_{i=1}^{N} (u_i - \hat{u}_i^{\text{text}})^2 = \sum_{i=1}^{N} \left( u_i - \sum_{k=1}^{n} \beta_{k,\text{text}} \cdot h_{i,k}^{\text{text}} \right)^2
\]

\subsection{Synthetic Utility Function, adapted from GNN \cite{Hoang2024GNNGN}}

In order to refine the model, we first estimate a synthetic utility function \( u_{\text{syn}}(\mathbf{x}) \) based on the combined coefficients \( \beta_{j,\text{num}} \) and \( \beta_{k,\text{text}} \):
\[
u_{\text{syn}}(\mathbf{x}) = \sum_{j=1}^{m} \beta_{j,\text{num}} \cdot x_j^{\text{num}} + \sum_{k=1}^{n} \beta_{k,\text{text}} \cdot h_k^{\text{text}}
\]

\subsection{Real Utility Function, adapted from GNN \cite{Hoang2024GNNGN}}

The final utility function \( u_{\text{real}}(\mathbf{x}) \) is then estimated using the synthetic utility function as shown below:
\[
u_{\text{real}}(\mathbf{x}) = \sum_{j=1}^{m} \gamma_{j,\text{num}} \cdot x_j^{\text{num}} + \sum_{k=1}^{n} \gamma_{k,\text{text}} \cdot h_k^{\text{text}} + \delta
\]
where \( \gamma_{j,\text{num}} \) and \( \gamma_{k,\text{text}} \) are the refined coefficients, and \( \delta \) is the error term.

\subsection{Final goal}

The main goal of GNN algorithm is to identify the optimal subsets \( t \subset T \) that maximize the real utility function \( u_{\text{real}}(\mathbf{x}) \):
\[
\text{Optimal Subsets} = \arg\max_{t \subset T} \sum_{i \in t} u_{\text{real}}(\mathbf{x}_i)
\]

\section{About the Indistinguishability query algorithm}
\begin{table}[h!]
\centering
\begin{tabular}{|c|c|c|c|}
\hline
\textbf{car} & \textbf{MPG} & \textbf{SR} & \textbf{MPG + 20SR} \\ \hline
$c_1$ & 59 & 5 & \cellcolor{gray!30}159 \\ \hline
$c_2$ & 36 & 4 & 116 \\ \hline
$c_3$ & 46 & 5 & \cellcolor{gray!30}164 \\ \hline
$c_4$ & 34 & 5 & 134 \\ \hline
$c_5$ & 35 & 5 & \cellcolor{gray!30}158 \\ \hline
\end{tabular}
\caption{Example of the indistinguishability query. All the highlighted tuples are 0.05-indistinguishable from the optimal ($c_3$) for a user with the given utility function.}
\end{table}

\noindent
\textbf{Example:} Consider Table I, as in Indistinguishibility Query \cite{Lall2024TheIQ}, which lists cars $c_1$ through $c_5$. Suppose Alice values fuel efficiency (MPG) and safety rating (SR) and has a utility function (unknown to her) given by $f(\text{MPG}, \text{SR}) = \text{MPG} + 20\text{SR}$. With this function, her top choice is $p^* = p_3$ since it achieves the highest utility score of 164. With an indistinguishability parameter of $\epsilon = 0.05$, Alice would be interested in any car that achieves at least $1/(1 + \epsilon) \approx 95.24\%$ of this maximum utility, which is approximately $156.2$ $(0.9524 \times 164)$. Therefore, the indistinguishability query should return cars $\{c_1, c_3, c_5\}$, as highlighted in the table. Note that, while $c_1$ differs significantly from Alice's ideal car $c_3$, it provides a similar utility value for her preferences.

\section{About the PLOD and GNN algorithm}

The workflow of the algorithm GNN: First, users rank the dataset's attributes, which are then scaled to the range $\{0,1\}$ based on the rankings. GNN estimates the coefficients for each attribute using machine learning for numerical data and GNN and LLM techniques for textual data. These coefficients are then converted into scores within the same range. The process continues by synthesizing an initial utility function based on these coefficients. GNN then refines this estimate to generate a final utility function, offering a robust approach to aligning predicted data subsets with user preferences.

\textbf{Graph Neural Networks and Large Language Models (GNN).} The GNN algorithm shows the strengths of both GNNs and LLMs, and PLOD handles datasets, including numerical and text features, by asking users to rank attributes and then leveraging GNNs and LLMs, and PLOD to estimate utility function coefficients. So, GNN offers a novel approach to data discovery that does not require explicit utility function definitions from the user. This method improves upon PLOD by incorporating advanced feature extraction and relational modeling capabilities, making it more robust to the complexities of multimodal data. Thus, GNN offers promising and reliance predicting data science and analytics applications.

\section*{Mathematical Proof for GNN algorithm}

By taking advantage of PLOD and Graph Neural Networks and Large Language Models, the GNN algorithm combines numerical and textual data processing using the advantage of PLOD as in \cite{hoang2024plod}, Graph Neural Networks (GNNs) as in several studies \cite{kipf2017semi} \cite{Hamilton2017InductiveRL} \cite{velickovic2018graph}, and Large Language Models (LLMs) as in \cite{devlin2019bert} \cite{brown2020language}, thus, to estimate the coefficients of a utility function that ranks the importance of different attributes to identify optimal subsets of data.

\section*{Utility Function and Error Term}

Adapted from GNN, the real utility function is defined as:
\[
u_{real}(x) = \sum_{j=1}^m y_{j,num} \cdot x_{j,num} + \sum_{k=1}^n y_{k,text} \cdot h_{k,text} + \delta
\]
where the below input can be denoted as:
\begin{itemize}
    \item \( m \): Number of numerical attributes.
    \item \( n \): Number of textual attributes.
    \item \( x_{j,num} \): Numerical attribute values.
    \item \( h_{k,text} \): Textual embeddings.
    \item \( y_{j,num} \), \( y_{k,text} \): Coefficients for numerical and textual attributes, respectively.
    \item \( \delta \): Error term.
\end{itemize}

The error term \( \delta \) is defined as the residual difference:
\[
\delta_i = u(x_i) - \left( \sum_{j=1}^m y_{j,num} \cdot x_{j,num,i} + \sum_{k=1}^n y_{k,text} \cdot h_{k,text,i} \right)
\]

\section*{Error Bound Derivation}

To derive the bound for \( \delta \), we consider contributions from numerical attributes, textual embeddings, and inherent noise.

\subsection*{Numerical Contribution}

For the numerical attributes, the contribution to the error is bounded by:
\[
\left| \sum_{j=1}^m y_{j,num} \cdot x_{j,num} \right| \leq \|y_{num}\|_1 \cdot \max(X_{num})
\]
where:
\begin{itemize}
    \item \( \|y_{num}\|_1 = \sum_{j=1}^m |y_{j,num}| \): The \( \ell_1 \)-norm of numerical coefficients.
    \item \( \max(X_{num}) \): Maximum value of numerical attributes.
\end{itemize}

\subsection*{Textual Contribution}

For the textual attributes, the contribution to the error is bounded by:
\[
\left| \sum_{k=1}^n y_{k,text} \cdot h_{k,text} \right| \leq \|y_{text}\|_1 \cdot \max(H_{text})
\]
where:
\begin{itemize}
    \item \( \|y_{text}\|_1 = \sum_{k=1}^n |y_{k,text}| \): The \( \ell_1 \)-norm of textual coefficients.
    \item \( \max(H_{text}) \): Maximum magnitude of textual embeddings.
\end{itemize}

\subsection*{Inherent Noise}

The inherent noise \( \epsilon \) accounts for the variability or randomness in the data that the model cannot explain. Thus, the error term includes a baseline noise component \( \epsilon \).

\section*{Final Error Bound}

Combining the contributions, the error term \( \delta \) is bounded by:
\[
|\delta| \leq \|y_{num}\|_1 \cdot \max(X_{num}) + \|y_{text}\|_1 \cdot \max(H_{text}) + \epsilon
\]

Expressing in terms of dataset parameters:
\begin{itemize}
    \item \( T = m + n \): Total number of attributes (numerical + textual).
    \item \( N \): Number of rows in the dataset.
    \item \( q \): Number of user-provided questions.
\end{itemize}

The error bound becomes:
\[
|\delta| \leq \frac{1}{\sqrt{N}} \left( m \cdot \max(X_{num}) + n \cdot \max(H_{text}) \right) + q \cdot \epsilon
\]

Thus, the bounds for the error of GNN \cite{Hoang2024GNNGN} is:

\[
0 \leq |\delta| \leq \frac{1}{\sqrt{N}} \left( m \cdot \max(X_{num}) + n \cdot \max(H_{text}) \right) + q \cdot \epsilon
\]

\begin{algorithm}
\caption{GNN: Graph Neural Networks and Large Language Models for Data Discovery \cite{Hoang2024GNNGN}}
\label{GNN Algorithm}
\begin{algorithmic}[1]  
\REQUIRE Dataset $D$ with features $X = \{x_1, x_2, \ldots, x_n\}$ (both numerical and text), and labels $y$
\ENSURE Subsets of optimal tuples $t \subseteq T$ with utility scores

\STATE \textbf{User Ranking:} Ask the user to rank all attributes $x_i \in X$
\FOR{each attribute $x_i \in X$}
    \STATE Scale down values of $x_i$ to the range $\{0,1\}$ based on ranking
\ENDFOR

\STATE \textbf{Coefficient Estimation:}
\STATE Train a Machine Learning model on numerical data to estimate coefficients $\beta_{num}$
\STATE Use Graph Neural Network (GNN) and Large Language Model (LLM) on text data to estimate coefficients $\beta_{text}$
\STATE Combine $\beta_{num}$ and $\beta_{text}$ to form overall coefficients $\beta = \{\beta_{num}, \beta_{text}\}$

\STATE \textbf{Synthetic Utility Function:}
\STATE Estimate synthetic utility function $U_{syn}$ based on coefficients $\beta$

\STATE \textbf{Real Utility Function:}
\STATE Refine and estimate real utility function $U_{real}$ coefficients based on $U_{syn}$

\STATE \textbf{Optimal Subsets Selection:}
\STATE Use $U_{real}$ to score and filter the dataset $D$ to find subsets of optimal tuples $t$

\STATE \textbf{Return} Optimal subsets $t$

\end{algorithmic}
\end{algorithm}

\section*{GNN: Graph Neural Networks and Large Language Models for Data Discovery\ \cite{Hoang2024GNNGN}}

In \textbf{lines 1-3:}, we are gonna let the GNN algorithm ask the user to rank all attributes in the dataset $X$, which contains numerical and textual data. Based on the user's ranking, the attributes are scaled down to a normalized range $\{0, 1\}$. In the \textbf{lines 4-8:}, We then try to estimate the coefficients for each attribute $\beta_{\text{num}}$ for numerical data. After this process, we then try to estimate the coefficients $\beta_{\text{text}}$ for textual data. Following, the algorithm then combines these estimated coefficients into a comprehensive set $\beta = \{\beta_{\text{num}}, \beta_{\text{text}}\}$ that shows the overall importance of each attribute. In the following \textbf{lines 9-10:} After that of using the combined coefficients, the algorithm tries to estimate a synthetic utility function $U_{\text{syn}}$, which is a step designed to approximate the actual utility function. Next in the \textbf{lines 11-12:}, GNN tries to refine this synthetic utility function to produce the actual utility function $U_{\text{real}}$. Next in \textbf{lines 13-14:} Finally, we then try with the real utility function, applied to score and filter the dataset $D$ to identify subsets of optimal tuples $t$. After all, in \textbf{line 15:} The algorithm finally returns the optimal subsets $t$ as the final output.

\section{Experiments}
The comparison of precision demonstrates the accuracy when all algorithms run on the same inputs that return how many retrieved results are relevant to the query. 

\section*{Why GNN Excels Over Other Algorithms}

\begin{table}[h!]
\centering
\begin{tabular}{@{}p{3cm}p{7cm}p{7cm}@{}}
\toprule
\textbf{Algorithm} & \textbf{Limitations}                                                                    & \textbf{Why GNN Excels}                                                                                  \\ \midrule

\textbf{PLOD}            & 
Fails to capture relationships between tuples and lacks semantic understanding of textual data. 
&
GNN integrates:
\begin{itemize}
    \item Numerical modeling using \( \text{SSE}_{num} \) for minimizing error in numerical predictions.
    \item Textual understanding using embeddings \( h_{k,text} \) generated by LLMs and refined by GNN.
    \item Relational learning through graph-based message passing.
\end{itemize} \\ \midrule

\textbf{Top-K}           & 
Assumes independence of attributes and ignores relationships between tuples. 
&
GNN dynamically adjusts rankings by:
\begin{itemize}
    \item Capturing relationships through graph propagation.
    \item Combining numerical and textual features in a unified synthetic utility function.
\end{itemize} \\ \midrule

\textbf{Skyline}         & 
Computationally expensive for high-dimensional datasets and lacks semantic understanding for textual data.
&
GNN efficiently handles high-dimensional data by:
\begin{itemize}
    \item Scaling computations using graph structures.
    \item Jointly optimizing numerical and textual attributes via the real utility function.
\end{itemize} \\ \midrule

\textbf{Multi-objective} & 
Requires manual weight tuning, leading to biases and inconsistencies in results. 
&
GNN optimizes holistically by:
\begin{itemize}
    \item Learning weights dynamically for both numerical and textual features.
    \item Identifying optimal subsets of tuples via the maximization of the real utility function.
\end{itemize} \\

\bottomrule
\end{tabular}
\caption{Why GNN Excels Compared to Other Algorithms}
\end{table}

\begin{figure}[H]
  \centering
  \includegraphics[width=0.6\linewidth]{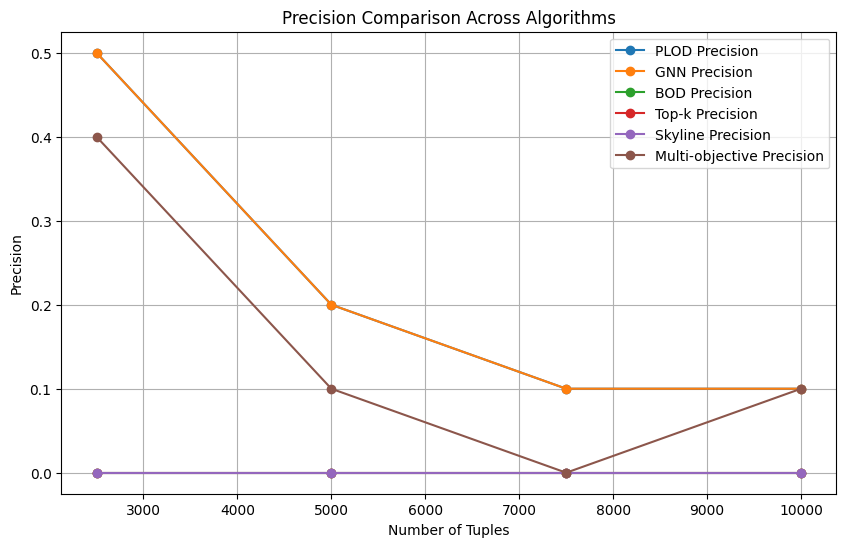} 
  \caption{Precision comparison with changes in number of tuples.}
  \label{fig:precision_comparison}
\end{figure}

\section{Conclusion}
The GNN performs with the highest precision compared to the previous approaches, as in Figure 1, but takes a comparable amount of time to accomplish. This happens because machine learning is used to estimate the coefficients compared to other approaches for numerical values, and graph neural networks and large language models are used to understand and weigh the textual values from a huge dataset, in which other approaches only understand the numerical values input from the user.

\bibliographystyle{ACM-Reference-Format}



\end{document}